# Dynamic reconstruction for atom probe tomography

Baptiste Gault[1,2], Shyeh Tjing Loi[1], Vicente J. Araullo-Peters[1], Leigh T. Stephenson[1], Michael P. Moody[1], Sachin L. Shrestha[1], Ross K.W. Marceau[1], Lan Yao[1], Julie M. Cairney[1], Simon P. Ringer[1]

[1]*Australian Centre for Microscopy and Microanalysis, Madsen Building F09, The University of Sydney, NSW 2006, Australia*

[2]*Institute of Materials and Engineering Science, Australian Nuclear Science and Technology Organisation, Private Mail Bag 1, Menai, NSW 2234, Australia*

## ABSTRACT

Progress in the reconstruction for atom probe tomography has been limited since the first implementation of the protocol proposed by Bas *et al.* in 1995. This approach, and those subsequently developed, assume that the geometric parameters used to build the three-dimensional atom map are constant over the course of an analysis. Here, we test this assumption within the analyses of low-alloyed materials. By building upon methods recently proposed to measure the tomographic reconstruction parameters, we demonstrate that this assumption can introduce significant limitations in the accuracy of the analysis. Moreover, we propose a strategy to alleviate this problem through the implementation of a new reconstruction algorithm that dynamically accommodates variations in the tomographic reconstruction parameters.

* Corresponding author: baptiste.gault@sydney.edu.au - Tel: + 61 2 93517548

# INTRODUCTION

Atom-probe tomography (APT) provides three-dimensional mapping of elements within a small volume of material with near-atomic resolution [1]. In APT, atoms are progressively desorbed and ionised from the surface of a needle-shaped specimen (Fig.1) under the effect of a very intense electric field ($\sim 10^{10}$ V.m$^{-1}$) in a process known as field evaporation [2]. The specimen has a radius of curvature ranging from 30 to 200 nm and is maintained at cryogenic temperatures (20-80 K). To enable elemental identification of each ion by time-of flight mass spectrometry, time-controlled field evaporation is induced by the combination of a DC voltage, of the order of a few kilovolts, and either high voltage (HV) [3] or fast (<10 ps) laser pulses [4, 5]. The intense, diverging electric field accelerates the ions away from the specimen surface and they are collected by a time-resolved position-sensitive detector [6, 7]. Using a relatively simple reverse-projection [8, 9] combined with an incremental increase of the depth, the impact coordinates of each ion on the detector is used to build a tomographic reconstruction of the field-evaporated volume within the field-of-view of the microscope [10, 11].

The original reconstruction protocol for the calculation of the lateral (*x,y*) coordinates for a specimen, which assumes that it is shaped as a spherical cap sitting on a truncated cone with shank angle $\alpha$ with tangential continuity between the cap and the cone (Fig.1), was developed together with the initial design of APT microscopes. At this time, the angular field-of-view of the experiment was limited to approximately 10 degrees [7] and hence small angle approximations were utilised to derive the equations that the protocol is based on. However, modern instruments now have an angular field-of-view of about 50 degrees [1, 12] and this approximation is no longer valid. Exact solutions for the point-projection were subsequently derived independently by several authors and implemented in commercial software for modern instruments [11, 13, 14]. The assumption of a simple point-projection enables a reduction of the number of unknown parameters to only two: the field factor, linking the amplitude of the electric field to the voltage and the specimen radius, and the image compression factor, accounting for the deformation of the ion trajectories toward the specimen axis (Fig.1). Different approaches have been proposed to compute the in-depth (*z*) coordinates of the atoms [10, 11, 14, 15]. All of these methods assume that the ions originate from a *virtual emitting surface* that progressively moves downwards by an increment proportional to the volume of each field-evaporated atom and is inversely proportional to a combination of the detection efficiency and a function $\frac{dv}{dz}$ that describes the increase in analyzed volume as the virtual emitting surface moves down [10, 11, 14]. Detection efficiency describes the fraction of ions within the field of view that are actually detected and is limited to about 55% on most microscopes due to the open area of the micro-channel plates. The values of the two geometrical reconstruction parameters are known to be connected to the geometry of the specimen and its electrostatic environment within the microscope [15, 16, 17, 18], and are hence specific to each experiment [18].

The intrinsic limitations of these protocols in cases where precipitates or multi-layer systems are imaged by APT have been extensively discussed in the literature [19, 20, 21, 22, 23, 24, 25]. This is partly related to the fact that, despite the knowledge that the reconstruction parameters evolve during the course of an experiment due to the progressive change in the specimen geometry [26, 27], they are considered by the reconstruction protocol to remain constant. Previously, values have often been reported for relatively short datasets or only for a section of the overall data [18], where variations in the parameters would be limited. However, considering the size of current routine APT analyses, the reconstruction parameters should be continually adjusted to reflect the evolution of the specimen geometry.

Even in the case of solid solutions or low-alloyed materials, where the accuracy of the tomographic reconstruction should be optimal, the protocol itself induces distortions, which ultimately affect the integrity of the data and hinder precise structural analysis [28]. In this article, we demonstrate how the methods introduced in refs. [18, 26] can efficiently be used to measure both the image compression and field factors at different points along the sequence of detection. Alongside enhanced protocols recently proposed by various authors [19, 28, 29], a new reconstruction protocol utilising dynamically adjusted reconstruction protocols is proposed. The validity of such an approach and the integrity of the resulting reconstruction are discussed.

## MATERIALS AND METHODS

### *EXPERIMENTAL*

APT analysis was undertaken on a series of Al-Cu-Mg alloys and low-alloy strip cast steels. Details on all of the specimens investigated in this study are reported in Table 1. Specimens were prepared for the atom probe by standard electropolishing methods using a solution of 25% perchloric acid in glacial acetic acid at 10 – 12 V DC, followed by a second stage of fine polishing under a binocular microscope using 2% perchloric in 2-butoxyethanol at 10 – 20 V DC. Specimens were analyzed using an Imago LEAP 3000X Si equipped with a delay line detector at a pulse fraction of 0.20-0.25. The specimens were maintained at cryogenic temperatures (<25 K) under ultrahigh vacuum conditions of $4.5 \cdot 10^{-9}$ Pa. A constant average detection rate of $0.5\text{-}2 \cdot 10^{-2}$ ions per pulse is ensured by controlling the total voltage applied to the tip. Datasets containing 12-100 million ions were considered.

### *TOMOGRAPHIC RECONSTRUCTION*

The tomographic reconstructions discussed here were built using the protocol described in ref. [11]. For each ion processed in the reconstruction, the specimen's current radius of curvature is calculated from the expression introduced by Gomer [30] to link the electric field to the voltage:

$$R = \frac{V}{k_f F_e} \quad (1),$$

where $k_f$ is the field factor and $F_e$ the strength of the electric field at the specimen apex. The strength of the electric field is assumed to be the evaporation field of the main constituent element of the material, the value of which can be derived from thermodynamic considerations and is tabulated in ref. [31]. In Fig. 1 the angle of inclination $\theta$ is defined, which approximately describes the direction normal to the surface of the specimen at the position from which the ejected ion is evaporated. The value of $\theta$ can be deduced from the compressed inclination angle $\theta'$ using $\theta = \theta' + \arcsin((\xi - 1)\sin\theta')$, where $\xi$ is the image compression factor. The value of $\theta'$ can be derived from the impact position on the detector: $\theta' = \arctan(D/L)$ with L the flight path and D the distance to the centre of the detector (Fig.1). Assuming knowledge of the radius of curvature, and that the azimuthal angle $\phi$ remains constant during the flight, it is possible to accurately determine the lateral coordinates of every ion's original position on the *virtual emitting surface*. The elevation of the virtual *emitting surface*, referred to as $z_{tip}$, is progressively moved downward for each processed ion by a small increment $dz$ proportional to a nominal atomic volume $\Omega$ of the detected species: $z_{tip}^{(i+1)} = z_{tip}^{(i)} + dz = z_{tip}^{(i)} + \frac{\Omega}{\eta S_A}$ where $S_A = 2\pi R^2 (1 - \cos\theta_D)$, which is the area of the spherical cap of radius R and delimited by the angle $\theta_D$, the angular half field-of-view. Measurement of the reconstruction parameters were performed in reconstructions built using the protocol developed by Geiser *et al*. [14]. These parameters were subsequently exploited to build reconstructions using the protocol described herein, yielding extremely similar results [11].

### CALIBRATION OF PARAMETERS

The image compression factor $\xi$ and field factor $k_f$ were measured using the partial crystallographic information present within the atom probe data. First, the image compression factor was estimated based on the relative positions of poles within the cumulative detector hit map, generally referred to as a *desorption map*. Poles, first observed in field ion microscopy, are features in the data that are directly related to major crystallographic directions present at the surface of the specimen [32]. They can be used for identifying the crystallographic orientation of the reconstructed specimen and further as a standard for comparing the imaged angles between crystallographic directions with the expected theoretical values to deduce the extent of angular compression [26, 33, 34, 35], enabling a direct measurement of $\xi$. In the second step, reconstructions were built using the measured $\xi$, and the value of $k_f$ was adjusted so as to obtain the correct crystallographic spacing between the atomic planes corresponding to a low-index crystallographic direction imaged within the dataset ({002}, {111} or {113} for Al datasets and {001} or {011} for the strip cast steels) [18]. It is worth noting that, despite probable variations in the electric field during the course of the analysis, only the product of the field factor and the electric field $k_f F_e$ appears within the reconstruction protocol. Here we assume that the evaporation field remained constant throughout the experiment, which, as discussed below, proves to be a valid assumption for the cases presented herein. Therefore, the calibration of the field factor accounts for potential changes in the electric field as well.

### *CRYSTALLOGRAPHIC MEASUREMENTS AND DATA QUALITY ASSESSMENT*

A variety of methods have been developed to interrogate how well reconstructed atom probe data can reproduce the crystallographic information within the original specimen. Spatial distribution maps (SDM) [36, 37, 38], Fourier transform (FT) approaches [39, 40], and more recently the three-dimensional Hough transformation (HT) [41] can all be used to determine both the interplanar spacing and the angles between crystallographic directions [42], enabling a precise assessment of the integrity of the tomographic reconstruction. In this study, all three methods were used to investigate the data. FT was used to estimate the interplanar spacing in the reconstruction calibration process. Advanced SDMs [38] were used to precisely estimate the interplanar spacing, while HT was exploited to derive the angular orientation of the main crystallographic direction with respect to the z-axis of the reconstruction, and subsequently determine the angle between crystallographic directions. These last two measurements were performed within successive slices of ~15 nm located every 20 nm in the reconstruction to estimate their change with increasing depth.

## RESULTS AND DISCUSSION

Each dataset was divided sequentially into a series of thin slices consisting of ~1 to 5 million ions within which the image compression factor was first measured. Subsequently, the field factor was deduced by adjusting the plane spacing of the crystallographic planes with the lowest Miller index, i.e. largest spacing, observed in the data. In turn, this enabled a precise estimation of the radius of curvature of the specimen. The values of both $\xi$ and $k_f F_e$ are reported as a function of the radius of curvature in Fig. 2 (a) and (b) respectively. Note that the value of the radius of curvature was derived from the measured total high-voltage applied to the specimen using eq. 1 and the electric field $F_e$ is taken as the tabulated value of the evaporation field for the pure element (Al or Fe) [43]. Additionally, the shank angle was extracted from a fit of the evolution of the high voltage as a function of the number of detected ions, and the ratio of the 1+ and 2+ charge states was measured. Exploiting the results of the post-ionization theory introduced by Haydock and Kingham [44], this ratio allowed for a precise estimation of the electric field amplitude. Variations in the measured electric field were consistently below 1% across all datasets and hence are not presented herein. These variations are accounted for in the calibration of $k_f$.

A consistent trend is observed whereby both the field and image compression factors decrease as the specimen progressively blunts during the analysis. This trend was expected based upon several previous theoretical studies [17, 28, 30] and some other preliminary measurements [27]. Importantly, if the parameters do vary independently, in turn the ratio $k_f F_e / \xi$ (which dictates the value of the interplanar spacing) also changes, as shown in Fig. 2(c). There is a clear relationship between specimen radius and this ratio, where increased radius correlates to a decreasing ratio. The overall specimen shape, in particular the shank angle and the ellipticity (flattening) of the apex, has been previously shown to influence the image compression and field factors [11, 16, 17]. However, here, we were not able to definitively correlate these parameters to the shank angle of the specimen, which is provided in Table 1, its composition or its orientation, (likely to affect the ellipticity of the specimen) [45]. Other parameters, such as the diameter of the aperture of the counter-electrode or the distance between the specimen and the electrode are likely to also play a role in the values and evolution of these geometrical factors.

Recent simulations developed by Vurpillot and co-workers [28] predicted that this ratio should be constant for a given specimen provided that its shank angle was constant. Further, they predicted a direct relationship between the image compression factor and field factor, whereby $\xi = k_f^{1/3}$. However, experimental evidence from the current study suggests that the simplistic view of the specimen modelled as a truncated cone with a unique and constant shank angle topped by a spherical cap is not suitable to describe the geometry of an actual atom probe specimen. Furthermore, an experimentally determined graph of $\xi$ as a function of $k_f$ shown in Fig. 3 demonstrates that this relationship does not accurately describe the behaviour of experimental data. Many parameters may contribute to create such a discrepancy, but it is likely that the some of the basic assumptions underpinning the simulations are not valid in the case of an actual experiment. For instance, due to the bias at the entrance of the detector, the ions do not fly in a field-free zone after the electrode as single-particle detectors incorporated in atom probe instruments use micro-channel plates (MCP) for ion-electron conversion and amplification, which, in the LEAP microscope, are biased to approximately - 3 kV. This high voltage tends to deflect the ion trajectories towards the detector. The extent to which the trajectories are affected depends on the voltage at which the atom has been field evaporated, but this bias results in an additional compression of the ion trajectories not related to the geometry of the specimen, which makes the ion trajectories harder to extrapolate. The intrinsic complexity of the actual electrostatic problem, from the nanoscale in the vicinity of the specimen surface to the macroscale at the detector, makes the use of such simulations almost impractical for the prediction of the evolution of the reconstruction parameters.

## NEW RECONSTRUCTION METHODOLOGY

There is no constraint in the reconstruction protocol introduced in ref. [11], and presented here so far, that requires the reconstruction parameters to remain constant during the reconstruction. In cases where there is enough crystallographic information available, such as the data presented here, then it is possible to improve the reconstruction protocol to account for potential variations in the parameters as the specimen is progressively analysed. The data in Fig. 2 can also be represented as a function of the sequence of detection. Hence, for each ion in the detection sequence a piecewise cubic interpolation function [46] can be used to determine a specific value of both the image compression and field factors as shown in Fig. 4(a). These functions were subsequently incorporated into an adapted version of the reconstruction procedure, replacing the respective static values, so as to dynamically account for the changes in the actual parameters. This new protocol is referred to as *dynamic reconstruction*. The difference in the volume of the resulting tomographic reconstruction in the case of the analysis of the 0.026% Nb hot-rolled strip cast steel that was heat treated for 30 s at 750 °C is highlighted in Fig. 4(b-c). Fig. 4(b) is the volume obtained using static reconstruction parameters calibrated within a slice of data approximately 20% of the way along the detection sequence. Fig. 4(c) is the reconstructed volume obtained via the dynamic reconstruction protocol. The improvement in the reconstruction accuracy can be assessed by the measurement of the interplanar spacing and the angle between crystallographic directions presented in Figures 5 and 6 respectively.

### *VALIDITY OF THE APPROACH*

Several datasets were reconstructed using this protocol. Their integrity was first investigated by comparing the change in the interplanar spacing of the main crystallographic planes observed in the dataset as a function of the detection sequence of the constituent atoms. An example of this is shown in Fig. 5 where the spacing of the (002) planes imaged in the analysis of the Al-2.18Cu-1.66Mg alloy is plotted as a function of the number of detected ions. In the case of the conventional reconstruction the value of the interspacing progressively drifts. However, when the dynamic reconstruction protocol is implemented the interspacing remains almost constant despite the change in the $k_f/\xi$ ratio. In a second analysis, the angles between different crystallographic directions were used as a metric to assess the reconstruction integrity. Using the 3D Hough transform method developed by Yao et al. [41], the orientation of 6 atomic plane families with respect to the analysis direction was determined. The angle between each pair-combination was calculated and then normalised by the corresponding known theoretical angle. The average ratio within each slice in the data is reported as a function of the analysis depth in Fig. 6. Again, a drift appears in the values observed in the reconstruction generated using the conventional reconstruction algorithm, while much more limited variations are encountered in the case of the dynamically reconstructed data. There is still some scattering around the ratio of 1, due to imperfections in the reconstruction that can be attributed to the presence of a large population of clusters in this dataset.

## SUMMARY AND CONCLUSION

In summary, we have demonstrated that the reconstruction parameters evolve during the course of an atom probe analysis. The evolution of both the image compression and field factor appear to be directly dependent on the increase in the radius of curvature, but were difficult to directly relate to either the specimen shank angle or crystallographic orientation. We have shown that, in analyses where clear crystallographic information can be extracted, the actual evolution of the parameters can be accounted for in the tomographic reconstruction protocol, which leads to much more accurate crystallographic reconstructions.

The results and method discussed herein represent a first step towards an enhanced tomographic reconstruction protocol. The dynamic reconstruction has been shown to be effective for these engineering solid solutions and low-alloy systems, even when there is a change in the specimen orientation within the same data set due to the presence of a grain boundary or when crystallographic defects (dislocations) are imaged. Further issues, such as local magnification [25, 47] due to the development of local curvatures at the specimen surface, which can be dramatic in some cases (such as for some multilayers for example [20]). Larson *et al.* recently demonstrated that the reconstruction of simulated data could be improved by an ad-hoc adjustment of the reconstruction parameters [19]. Even in these more complex cases, the evolution of the reconstruction parameters with the global specimen shape will have to be accounted for by the protocol.

Ideally, the reconstruction parameters could be predicted using simulations, so as to enable the application of this protocol to data in which less crystallographic information is retained, as this requirement currently limits the applicability of the methods presented herein. Considering that the trends extracted from electrostatic simulations presented by Vurpillot and co-workers [28] do not always reflect experimental measurements, it is likely that the are many factors still not accounted for that influence the evolution of the reconstruction parameters, making their prediction non-trivial. This indicates that only sophisticated and very accurate simulations of the whole microscope setup will enable a full prediction of their evolution. Other parameters are likely to vary and could also be measured. For example, the change in the analysed area due to the distortion towards the detector at relatively low voltages was not accounted for here. Despite not enabling to correct for these additional issues, which represent possible future refinements, our present work represents a much-needed improvement on the previously proposed reconstruction protocols, forming a basis for more evolved reconstruction methods.

## ACKNOWLEDGEMENTS

The authors are grateful for scientific and technical input and support from the Australian Microscopy & Microanalysis Research Facility (AMMRF) at The University of Sydney. Drs Brian Geiser and David Larson are sincerely thanked for fruitful discussions and positive interactions. Mr G. da Costa and Dr F. Vurpillot (GPM, Université de Rouen – CNRS, France) are acknowledged for provision of the Fourier transform calculation software.

## FIGURE AND TABLE CAPTIONS

Figure 1: Schematic view of the atom probe microscope.

Table 1: Description of the different specimens analysed.

Figure 2: Evolution of (a) the image compression factor $\xi$, (b) the product of the field factor and the tabulated evaporation field $k_f F_e$, and (c) the ratio $k_f F_e / \xi$ as a function of the specimen radius of curvature for a series of different specimens (see Table 1 for legend). The colours and symbols are consistent across all subsequent figures.

Figure 3: Plot of the image compression factor as a function of the field factor for all specimens investigated (see Table 1 for legend). The dashed line corresponds to the evolution $\xi = k_f^{1/3}$ predicted by simulations.

Figure 4: (a) Determination of $\xi$ and $k_f$ for each ion by means of a piecewise cubic interpolation function. Reconstructions obtained using either the conventional (b) or the dynamic (c) protocol. The change in the shape and dimensions of the overall reconstruction are readily visible. Note that only the C atoms and a small fraction (<1%) of the Fe ions are displayed.

Figure 5: Interplanar spacing and $k_f/\xi$ ratio as a function of the number of detected ions for the Al-2.18Cu-1.66Mg alloy data.

Figure 6: Average ratio between the crystallographic and measured angles between atomic plane families as a function of the depth within the tomographic reconstruction built using the conventional (blue) or the dynamic (red) protocol. The data used was from the strip steel presented in Fig.4.

| Symbol | Alloy (wt. %) | Thermomechanical treatment | Shank Angle |
|---|---|---|---|
| 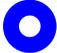 | Al-1Cu-0.76Mg | 1 h at 525 ºC - water quenched (WQ) | 5.7 ° |
| 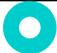 | Al-3Cu-2.28Mg | 1 h at 525 ºC – WQ | 9.5 ° |
| 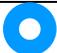 | Al-3Cu-2.28Mg | 1 h at 525 ºC – WQ – 3 months natural ageing | 16.7 ° |
| 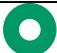 | Al-1Cu-0.76Mg | 1 h at 525 ºC – WQ | 16.2 ° |
| 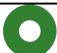 | Al-2.18Cu-1.66 | 1 h at 525 ºC – WQ - 300 s at 150 °C – QW | 13.7 ° |
| 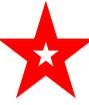 | 0.24Si-0.006N-0.031C-0.83Mn 0.084Nb-Bal. Fe | Hot rolled | 4.2 ° |
| 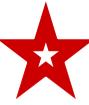 | 0.24Si-0.006N-0.031C-0.83Mn 0.084Nb-Bal. Fe | Hot rolled | 3.6 ° |
| 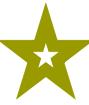 | 0.28Si-0.005N-0.037C-0.93Mn 0.065Nb-Bal. Fe (wt. %) | Hot rolled | 4.8 ° |
| 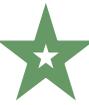 | 0.28Si-0.005N-0.037C-0.93Mn 0.065Nb-Bal. Fe | Hot rolled | 9.7 ° |
| 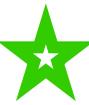 | 0.24Si-0.005N-0.038C-0.87Mn 0.026Nb-Bal. Fe | Hot rolled | 4.7 ° |
| 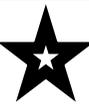 | 0.21Si-0.007N-0.036C-0.87Mn 0.041Nb-Bal. Fe | Hot rolled – 30 sec at 750°C | 4.4 ° |

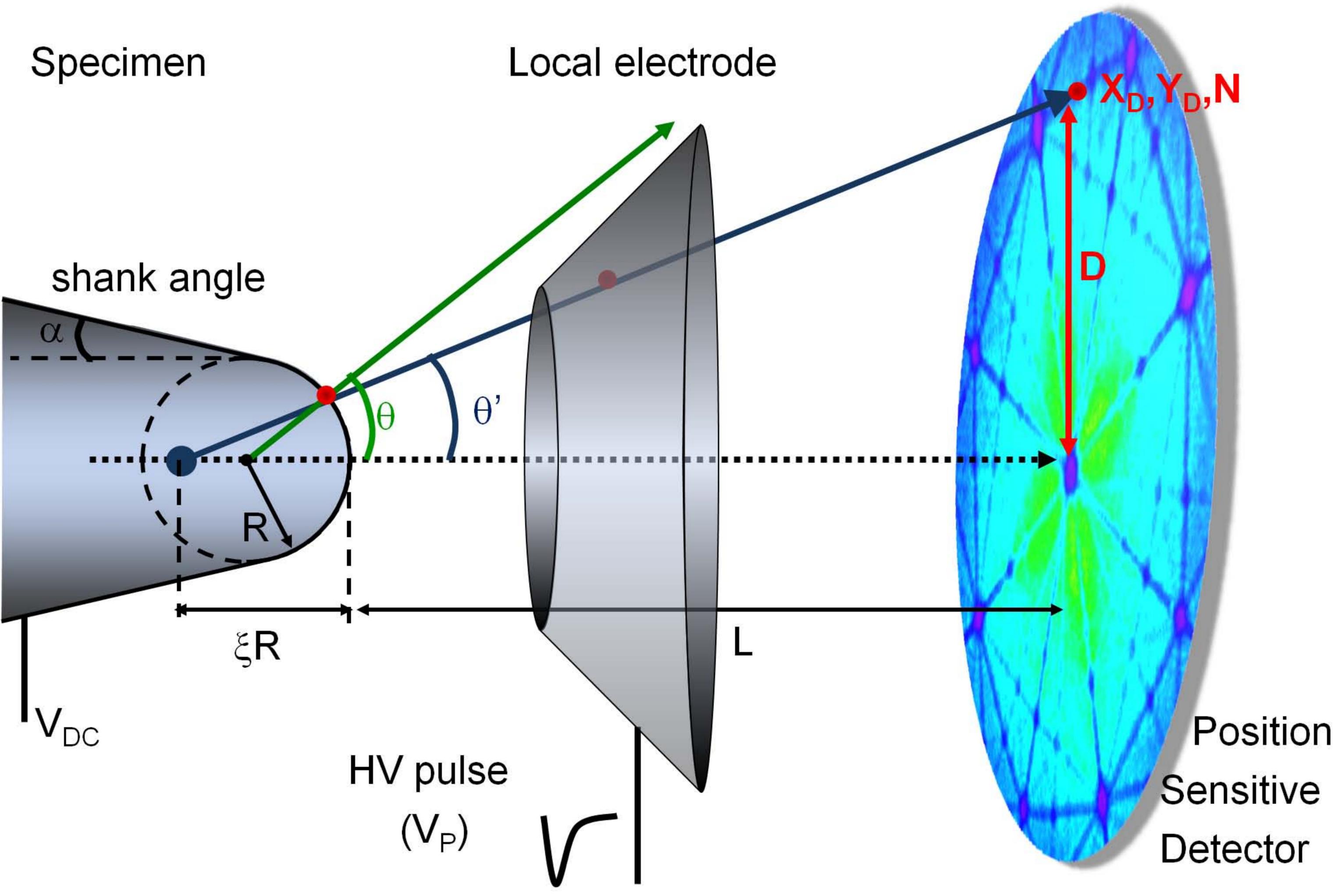

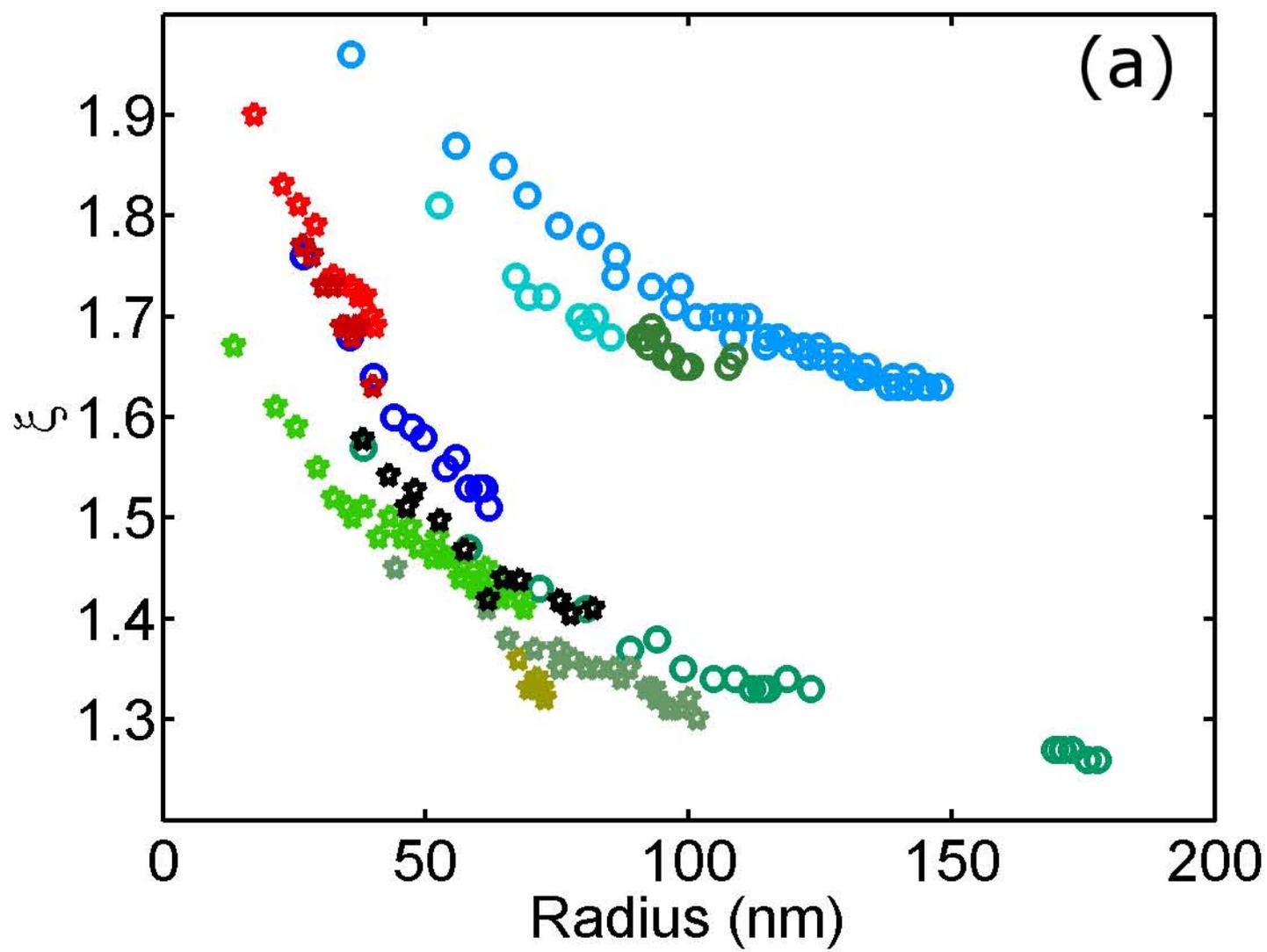
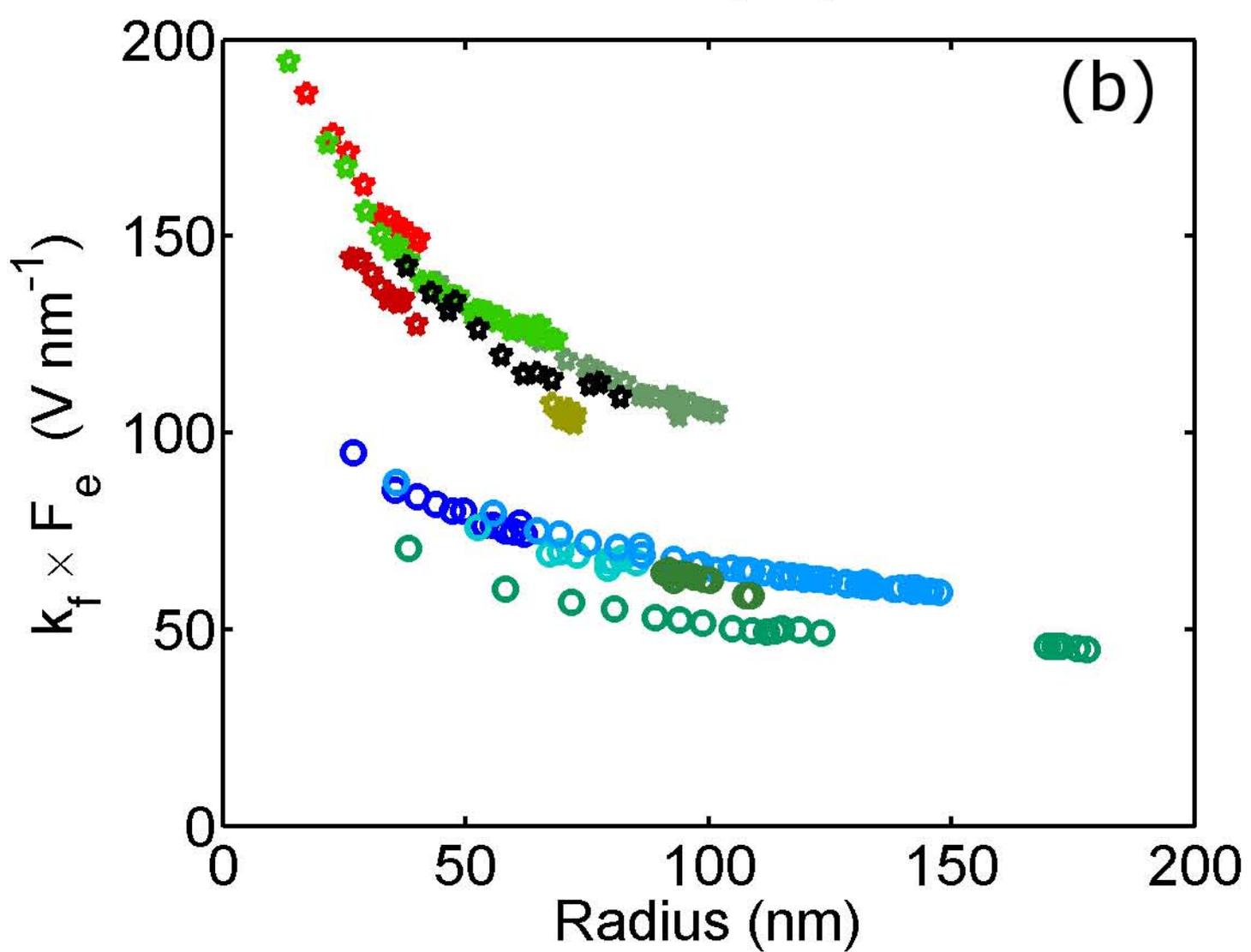
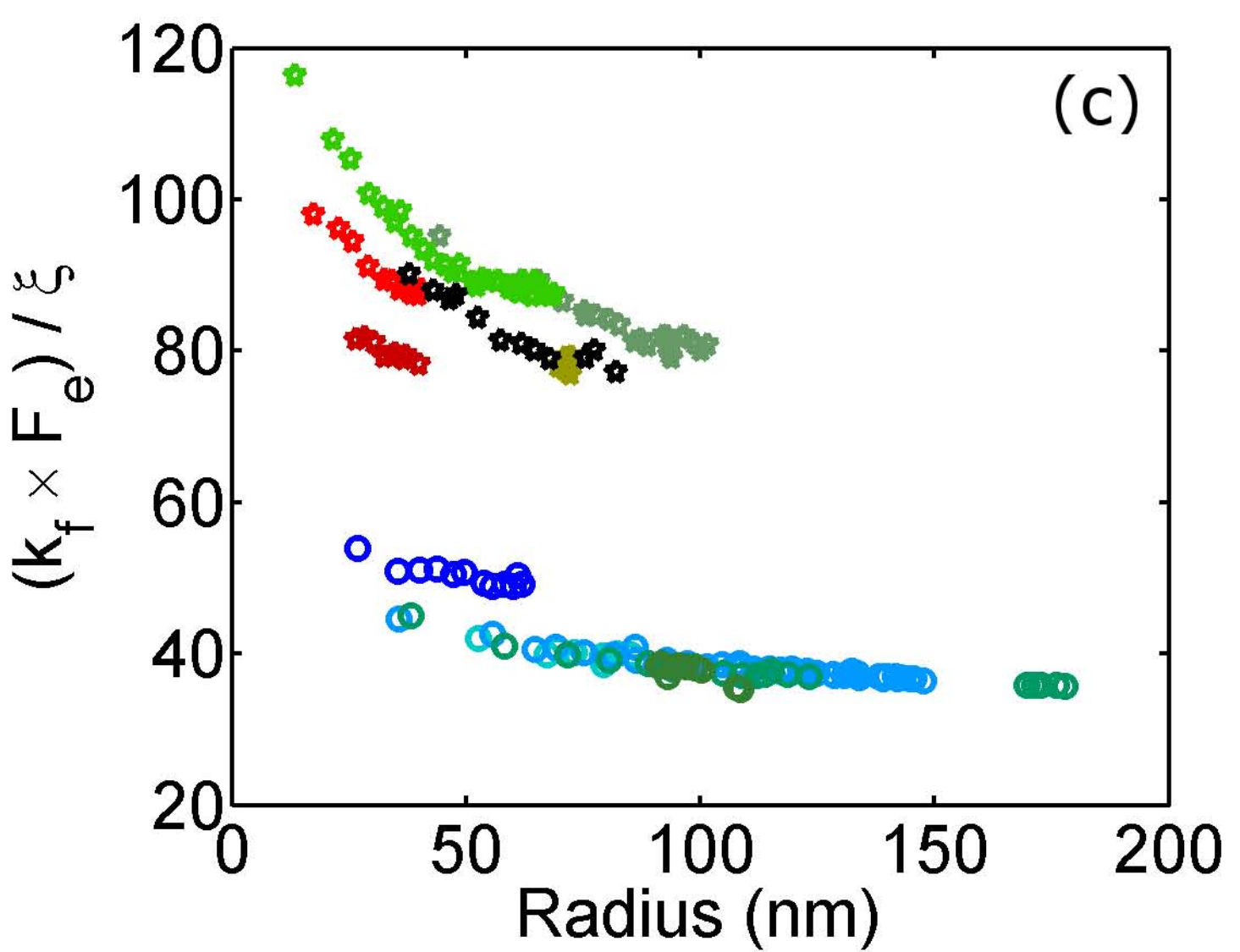

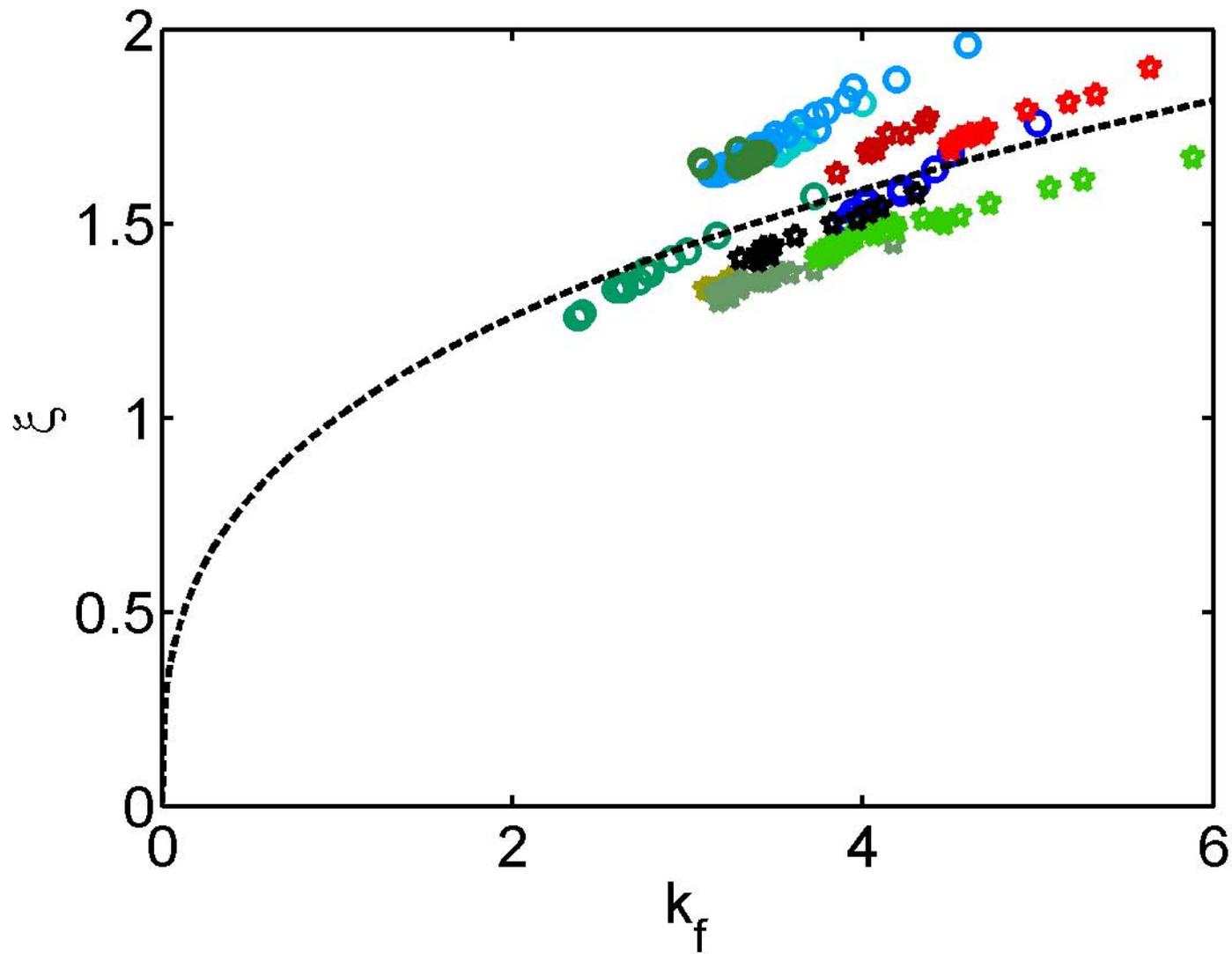

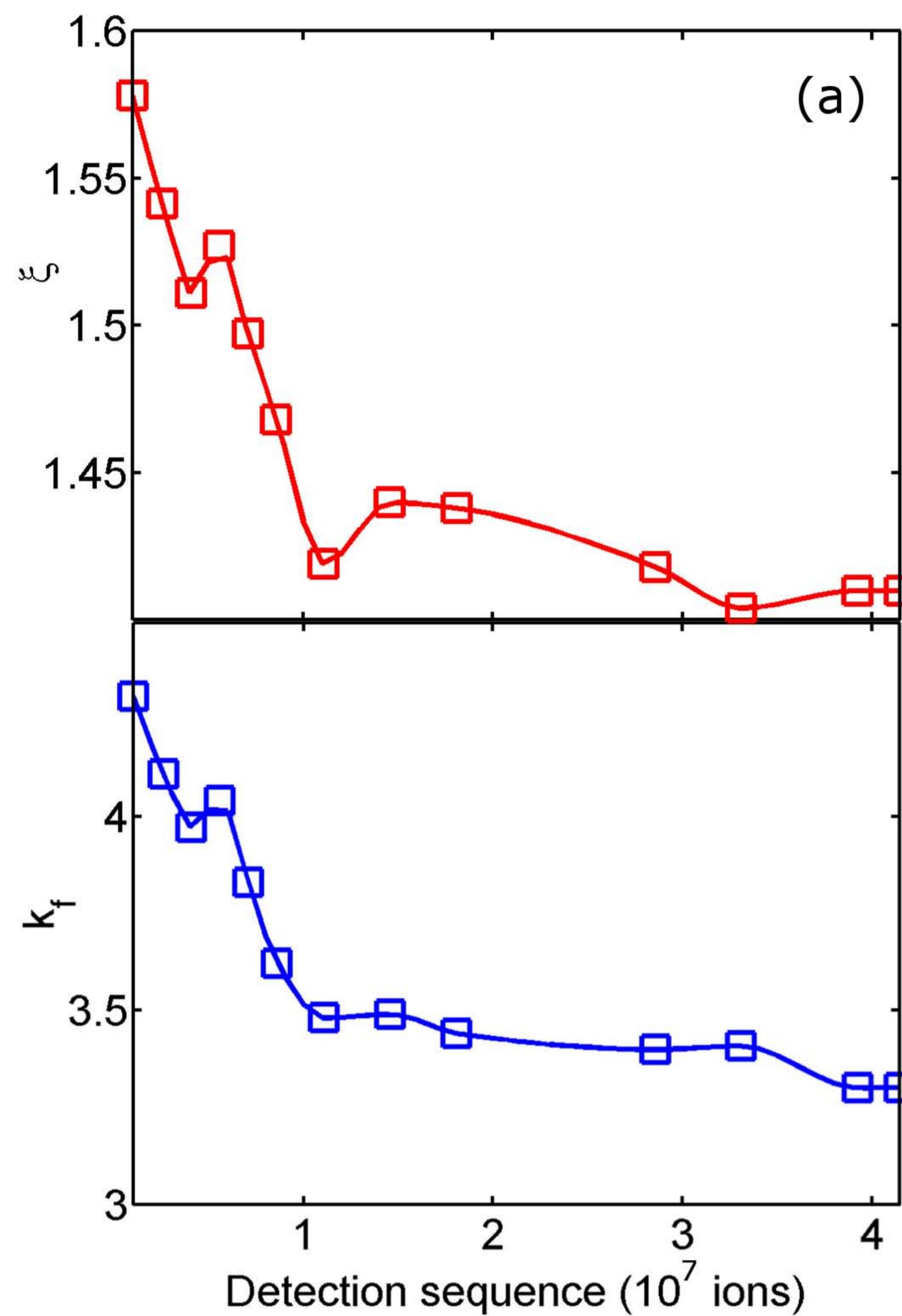
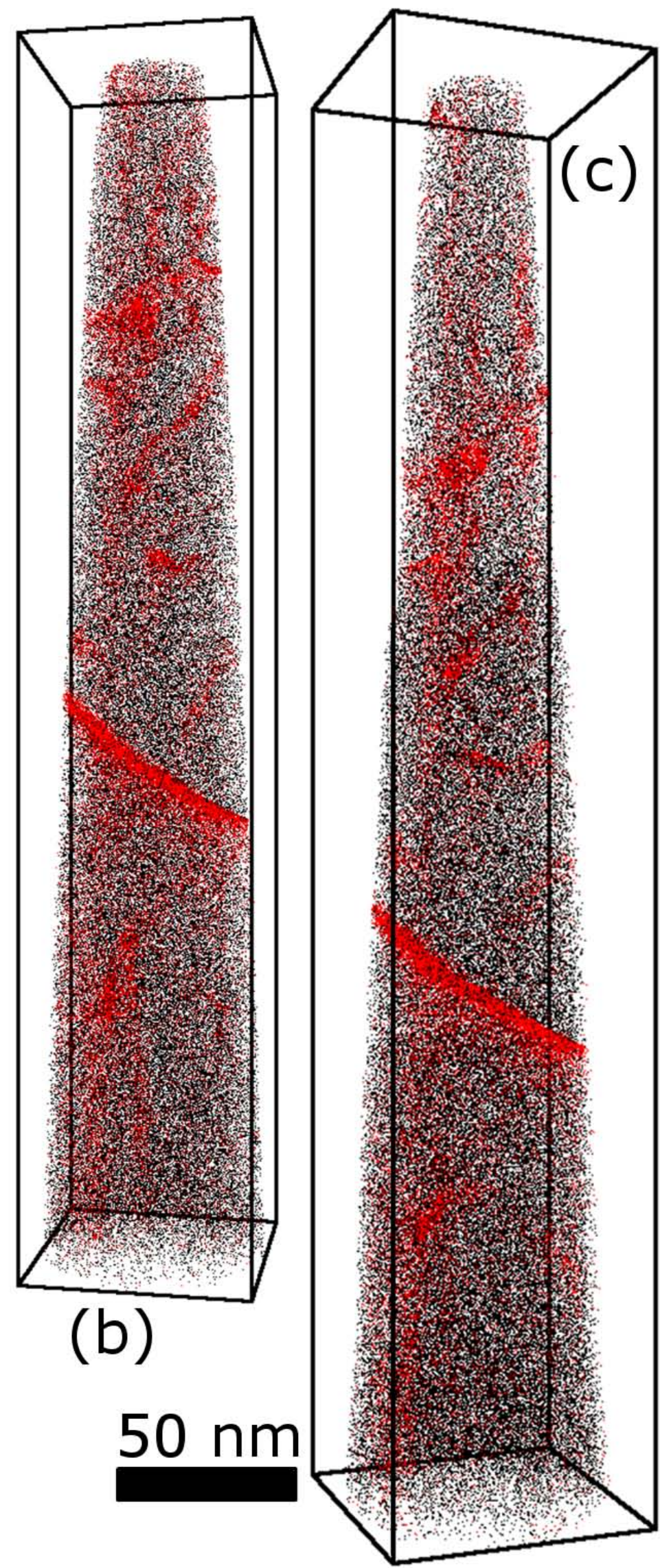

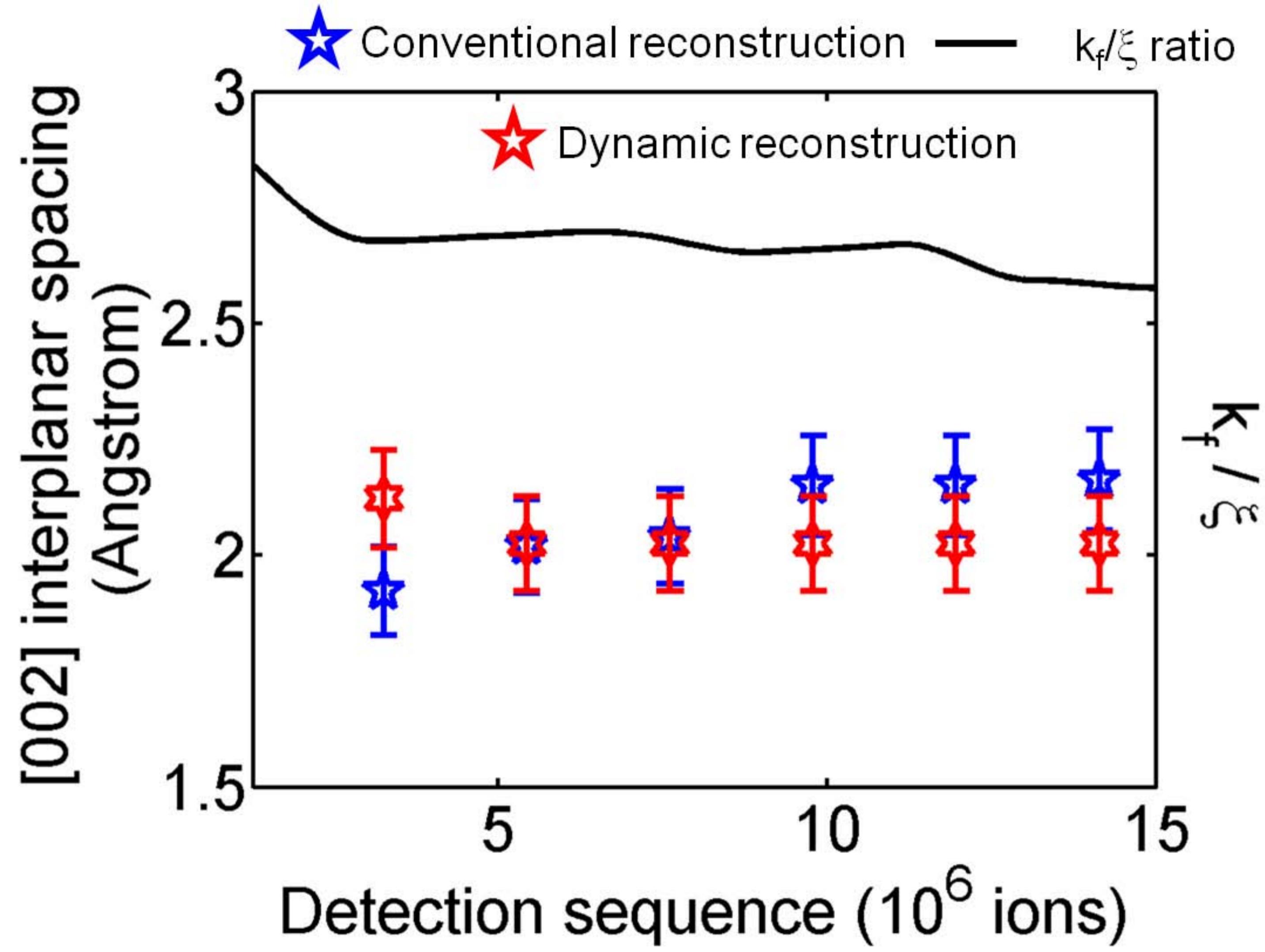

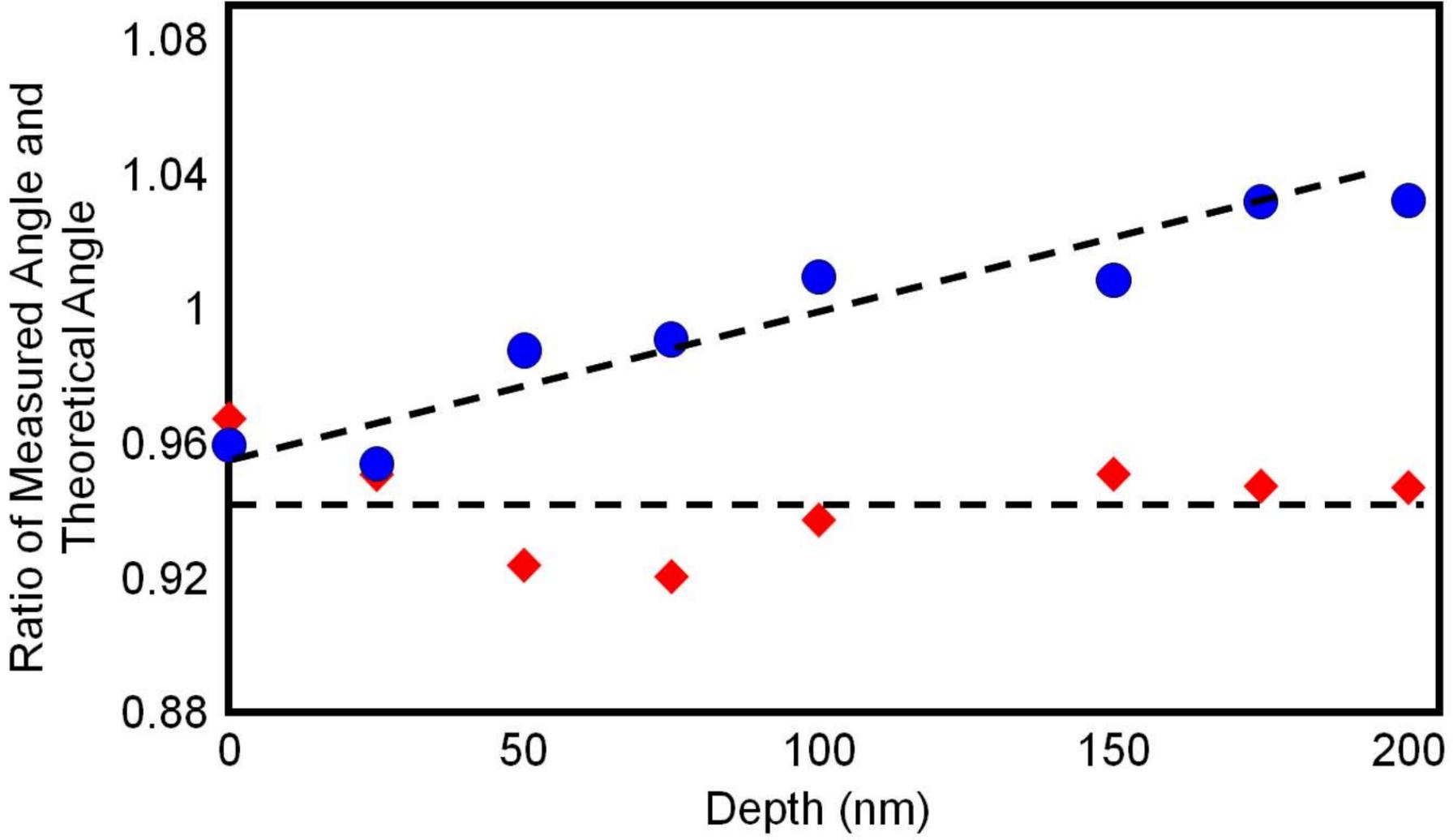